\documentstyle[epsf,tighten,preprint,pre,aps]{revtex}
\begin{document}
% uncomment next line for two-column draft mode
%\twocolumn[\hsize\textwidth\columnwidth\hsize\csname @twocolumnfalse\endcsname
\draft
\title{Field-Dependent Tilt and Birefringence of Electroclinic Liquid Crystals:
Theory and Experiment}
\author{Jonathan V. Selinger,$^1$ Peter J. Collings,$^{1,2}$ and R.
Shashidhar$^1$}
\address{$^1$Center for Bio/Molecular Science and Engineering,
Naval Research Laboratory, Code 6900, \\
4555 Overlook Avenue, SW, Washington, DC  20375 \\
$^2$Department of Physics and Astronomy, Swarthmore College, Swarthmore, PA
19081}
\date{September 4, 2001}
\maketitle

\begin{abstract}
An unresolved issue in the theory of liquid crystals is the molecular basis of
the electroclinic effect in the smectic-A phase.  Recent x-ray scattering
experiments suggest that, in a class of siloxane-containing liquid crystals, an
electric field changes a state of disordered molecular tilt in random
directions into a state of ordered tilt in one direction.  To investigate this
issue, we measure the optical tilt and birefringence of these liquid crystals
as functions of field and temperature, and we develop a theory for the
distribution of molecular orientations under a field.  Comparison of theory and
experiment confirms that these materials have a disordered distribution of
molecular tilt directions that is aligned by an electric field, giving a large
electroclinic effect.  It also shows that the effective dipole moment, a key
parameter in the theory, scales as a power law near the smectic-A--smectic-C
transition.
\end{abstract}

\pacs{PACS numbers:  61.30.Cz, 61.30.Gd, 64.70.Md}

% uncomment next line for two-column draft mode
%\vskip2pc]
\narrowtext

\section{Introduction}

In liquid crystals, an applied electric field influences the orientational
order of the molecules.  In particular, in the smectic-A (SmA) phase of chiral
liquid crystals, an electric field applied in the smectic layer plane induces a
molecular tilt relative to the layer normal.  The magnitude of the tilt varies
continuously with electric field, and the direction of the tilt is orthogonal
to the field.  This coupling between an electric field and the molecular tilt
is called the electroclinic effect.  It was predicted on the basis of
symmetry~\cite{meyer77} and was subsequently observed
experimentally~\cite{garoff77}.  It is now being developed for use in
electro-optic devices in which the continuously variable tilt leads to a gray
scale~\cite{collings89,andersson89,williams91,ratna93}.

Most theoretical understanding of the electroclinic effect has been developed
through Landau theory, which minimizes the free energy expanded in powers of
the molecular tilt and polarization~\cite{meyer77,carlsson88,abdulhalim91}.
This phenomenological approach explains macroscopic aspects of the
electroclinic effect.  It shows that the tilt depends linearly on electric
field for low fields, and that the coefficient of the linear dependence
diverges as the system approaches a second-order transition from the SmA to the
smectic-C (SmC) phase.  However, the microscopic basis of the electroclinic
effect is still unresolved.  Key questions are: What is the distribution of
molecular orientations, and how does this distribution change under an applied
electric field?

There have been two general concepts about the microscopic basis of the
electroclinic effect.  In the first scenario, the molecules all stand
perpendicular to the smectic layers in the absence of a field, and they
reorient together as rigid rods under a field.  In the second scenario, the
molecules have a random distribution of azimuthal orientations about a tilt
cone before the field is applied, and they become ordered in a single tilted
direction under a field.  The latter scenario is suggested by the de Vries
description of the SmA phase~\cite{devries79}.  Each of these concepts is
consistent with a net observed tilt that scales linearly with applied electric
field for low fields, and then saturates at high fields.  Thus, the issue is
how to distinguish between these possibilities.

One way to distinguish between these microscopic scenarios is through
molecular-scale simulations.  Our group has carried out Monte Carlo simulations
of SmA liquid crystals under an applied electric field~\cite{xu99}.  These
simulations use a model molecular structure consisting of seven soft spheres
bonded rigidly together in the biaxial zig-zag shape of the letter Z.  A
transverse electric dipole moment makes the molecules chiral.  These
simulations show a strong electroclinic effect, which involves a combination of
the ``rigid-rod'' and ``de Vries'' scenarios.  In the absence of an electric
field, the molecules have a distribution of orientations, with vortex defects
in the smectic layers.  When an electric field is applied, the magnitude of the
tilt increases {\it and\/} the azimuthal orientation of the tilt becomes
ordered, perpendicular to the electric field.  Thus, the simulations show that
both of these scenarios can occur in model liquid crystals.  They do not,
however, show which of these scenarios plays the dominant role in actual
experimental materials.

To distinguish between these possibilities in experimental materials, several
studies have examined the smectic layer spacing as a function of applied
electric field.  The two scenarios make very different predictions for the
smectic layer spacing.  In the rigid-rod scenario, when the molecules tilt by
an angle $\theta$, the smectic layer spacing should contract by a factor of
$\cos\theta$.  By contrast, in the de Vries scenario, the molecules are already
tilted in zero field, and a field only orders the azimuthal direction of the
tilt, so the field should not induce any layer contraction.  The experimental
studies have found that most ``conventional'' SmA liquid crystals show a
field-induced layer contraction, consistent with the prediction of the
rigid-rod scenario.  This contraction can be seen in measurements of the layer
spacing through x-ray diffraction~\cite{crawford94}.  It can also be seen
through field-induced layer buckling, which gives an optical stripe
pattern~\cite{crawford94,pavel91,skarp92,geer98,bartoli98}.  However, certain
materials have been developed that show a substantial electroclinic tilt with
hardly any layer contraction, consistent with the de Vries scenario.  These
include compounds with a fluoroether tail~\cite{radcliffe99}, a chiral lactic
ester in the tail~\cite{giesselmann99}, and dimethylsiloxane groups in the
tail~\cite{naciri95}.  In fact, an optical and x-ray study of one
organosiloxane compound revealed tilt angles as large as 31$^\circ$ in the
SmA phase with a layer contraction of less than 1\%~\cite{spector00}.

The purpose of our current study is to explore a different way of
distinguishing between these possibilities.  Instead of measuring the smectic
layer spacing, we investigate the optical birefringence as a function of
applied electric field.  The birefringence is the difference between indices of
refraction for light that is linearly polarized parallel or perpendicular to
the average director of a sample.  It is an appropriate probe for the
microscopic basis of the electroclinic effect because it is sensitive to the
degree of orientational order.  The rigid-rod and de Vries scenarios make
different predictions for the birefringence as a function of electric field.
In the rigid-rod scenario, the molecules have strong orientational order even
in zero field, so the zero-field birefringence should be high.  When an
electric field is applied, the molecules remain parallel to each other in a
tilted orientation, and hence the birefringence should vary only weakly as
function of field~\cite{bartoli97}.  By contrast, in the de Vries scenario, the
molecules have a distribution of orientations about a tilt cone in zero field.
The zero-field birefringence should be greatly reduced because of the
orientational averaging about the tilt cone.  When an electric field is
applied, the molecules become more aligned with each other in a particular
tilted orientation.  As a result, the birefringence of a de Vries-type material
should increase substantially with applied field.

In a preliminary communication, our group reported experiments on the optical
tilt and birefringence of four electroclinic liquid crystals with closely
related chemical structures:  KN125, SiKN105, DSiKN65, and
TSiKN105~\cite{lindle99}.  In these abbreviations, KN is a label, the numbers
on the right refer to the length of the hydrocarbon chains, and the letters on
the left refer to siloxane units in the latter three compounds.  KN125 is
believed to follow the rigid-rod scenario for the electroclinic effect (based
on a substantial layer contraction and buckling~\cite{crawford94}), while the
three siloxane-containing compounds are believed to follow the de Vries
scenario (based on the lack of layer buckling).  Our experiments confirmed that
KN125 has a large and weakly field-dependent birefringence, while the
siloxane-containing compounds have a much smaller and more strongly
field-dependent birefringence.  To analyze the data, we developed a model for
the birefringence as a function of field in de Vries-type materials, based on
averaging the molecular dielectric tensor over a field-dependent orientational
distribution function.  This model was consistent with the observed
field-dependent birefringence in the siloxane-containing compounds.

In this paper, we go beyond that preliminary communication to present a
detailed theoretical and experimental study of the optical tilt and
birefringence in two of the siloxane-containing liquid crystals, DSiKN65 and
TSiKN65.  On the experimental side, we measure the tilt and birefringence as
functions of temperature as well as applied electric field.  These measurements
show that the tilt and birefringence depend sensitively on temperature near the
SmA-SmC phase transition.  On the theoretical side, we develop a systematic
model for the orientational distribution in de Vries-type materials through a
series of manipulations of the dielectric tensor, and we note that this model
predicts the optical tilt as well as the birefringence.  Hence, we use the
model to fit the ensemble of data for tilt and birefringence as functions of
field and temperature.  The overall quality of the fits is fairly good,
considering that a simple model is being applied to a large data set.  For that
reason, we can conclude that the model captures the essential features of the
orientational ordering in de Vries-type materials.  Furthermore, comparison
between theory and experiment allows us to extract an important theoretical
parameter, the effective dipole moment, as a function of temperature.  We find
that this quantity scales as a power law near the SmA-SmC transition.  That
scaling is consistent with predictions from the theory of critical phenomena.

The outline of this paper is as follows.  In Sec.~II we present the
experimental method and results, showing the dependence of optical tilt and
birefringence on both electric field and temperature.  In Sec.~III we develop
the theory for the orientational ordering in de Vries-type materials, leading
to predictions for optical tilt and birefringence.  We compare the theory with
the experiment in Sec.~IV, in order to assess the quality of the fit and
extract the effective dipole moment.  In Sec.~V we discuss the results and
present the overall conclusions of this theoretical and experimental work.

\section{Experiment}

The two siloxane-containing compounds used in this investigation, DSiKN65 and
TSiKN65, have the structure and transition temperatures shown in Fig.~1.  These
liquid crystals were vacuum-filled into EHC cells of 5~$\mu$m thickness with
rubbed polyimide surfaces. The bookshelf geometry of the SmA phase was achieved
by extremely slow cooling through the isotropic-SmA transition in the presence
of a 1~Hz bipolar square-wave electric field with an amplitude of 5~V/$\mu$m.
The temperature of the sample cell was regulated by an Instec mK-2 controller
and HS-1 hotstage.  The temperature gradient across the portion of the sample
being illuminated was less than 0.1~K. The hotstage was placed on the rotable
stage of a polarizing microscope with a $\times10$ eyepiece and $\times5$
objective.  The light from a halogen lamp passed through a 633~nm filter
(${\rm FWHM}=3$~nm) before encountering the sample.  The intensity of the
transmitted light was measured by a silicon diode detector, amplifier, and
oscilloscope.  At each temperature, various electric-field values were applied
to the sample by a bipolar 10 Hz square wave.

For a homogeneous liquid crystal sample between crossed polarizers, with its
director perpendicular to the light propagation direction, the transmitted
intensity $I_\perp(\gamma)$ is given by
\begin{equation}
I_{\perp}(\gamma) = I_{\rm min} + I_{0} \sin^{2}(\delta/2) \sin^{2}(2\gamma),
\end{equation}
where $I_{\rm min}$ is the background intensity, $I_{0}$ is the incident
intensity, $\gamma$ is the angle between the director and either of the
polarizer axes, and $\delta$ is the phase retardation angle.  The latter angle
depends on the birefringence $\Delta n$, the sample thickness $d$, and the
wavelength of light $\lambda$ through
\begin{equation}
\delta = 2 \pi d \Delta n/\lambda.
\end{equation}
As the sample stage is rotated, the maximum value
$[I_{\perp}(\gamma)]_{\rm max}$ occurs when $\gamma = \pi/4$, and the minimum
value $I_{\rm min}$ occurs when $\gamma = 0$.

If the polarizers are parallel to each other instead of crossed, the
transmitted light intensity $I_{\parallel}(\gamma)$ is
\begin{equation}
I_{\parallel}(\gamma) = I_{\rm min} + I_{0} [1 - \sin^{2}(\delta/2)
\sin^{2}(2\gamma)].
\end{equation}
Rotation of the sample stage yields the maximum value,
$[I_{\parallel}(\gamma)]_{\rm max}= I_{\rm min} + I_{0}$ at $\gamma = 0$, and
the minimum value $[I_{\parallel}(\gamma)]_{\rm min}$ when $\gamma = \pi/4$.
Measurement of the minimum and maximum values of the intensity with the two
polarizer configurations in place can be used to
find the phase retardation angle,
\begin{equation}
\delta = 2 \sin^{-1}
\sqrt{\frac{[I_{\perp}(\gamma)]_{\rm max}-I_{\rm min}}
{[I_{\parallel}(\gamma)]_{\rm max}-I_{\rm min}}},
\end{equation}
and hence the birefringence $\Delta n$.

The tilt angle can be easily measured by rotating the sample stage so that
$I_\perp(\gamma)$, the transmitted intensity with the polarizers crossed, is
equal for both halves of the bipolar square wave.  In the two halves of the
square wave, the director orientation is $\gamma=\gamma_0\pm\theta_{\rm tilt}$,
where $\gamma_0$ is the orientation of the layer normal relative to either
polarizer axis and $\theta_{\rm tilt}$ is the electroclinic tilt angle.  If the
intensities are equal, then $\gamma_0 = 0$ and hence the intensity
$[I_{\perp}]_{\pm}$ is just
\begin{equation}
[I_{\perp}]_{\pm} = I_{\rm min} + I_{0} \sin^{2}(\delta/2)
\sin^{2}(2\theta_{tilt}).
\end{equation}
If this measurement is combined with the measurements of the maximum and
minimum intensities with crossed polarizers, the tilt angle can be determined
as
\begin{equation}
\theta_{\rm tilt} = \frac{1}{2} \sin^{-1}
\sqrt{\frac{[I_{\perp}]_{\pm}-I_{\rm min}}
{[I_{\perp}(\gamma)]_{\rm max}-I_{\rm min}}}.
\end{equation}

We measured the tilt angle and birefringence for eleven values of the electric
field at eight values of the temperature, starting just above the SmC-SmA
transition and ending roughly 10~K above the transition.  The data for DSiKN65
are shown by the symbols in Figs.~2(a) and~2(b), and the data for TSiKN65 are
shown in Figs.~3(a) and~3(b).  Several features of the data are clear from
these figures.  The tilt angle increases linearly with electric field at low
field and then saturates at an asymptotic value at high field.  The
birefringence increases quadratically with field at low field and then
saturates.  Both of these quantities depend more sensitively on field near the
SmC-SmA transition temperature than at higher temperature, away from the
transition.  By comparison with ``conventional'' electroclinic liquid crystals
that follow the rigid-rod scenario, such as KN125~\cite{bartoli97}, the
birefringence of these materials is much smaller and varies much more with
electric field.

For an alternative way to look at the data, we plot the birefringence vs.\ tilt
angle for DSiKN65 and TSiKN65 in Figs.~2(c) and~3(c), respectively.  The most
striking feature of these plots is that, for each material, the measurements at
all temperatures collapse onto a single universal curve.  The shape of this
curve is approximately a parabola.

\section{Theory}

To explain the dependence of the birefringence and tilt of DSiKN65 and TSiKN65
on electric field and temperature, we develop a theory for orientational
ordering in de Vries-type materials.  This theory is related to the theory for
field-induced biaxiality in ``conventional'' rigid-rod SmA liquid
crystals~\cite{bartoli97}.  It is also similar to the ``random model'' for the
optical properties of V-shaped switching materials~\cite{inui96,park99}.  One
difference from the latter is that it takes into account the inherent
biaxiality of the molecules.

This theory is based on a rotational averaging of the dielectric tensor
$\epsilon$.  In the coordinate system of a single molecule, the dielectric
tensor at optical frequencies has the diagonal form
\begin{equation}
\epsilon=\left(
\begin{array}{ccc}
\epsilon_a & 0 & 0 \\
0 & \epsilon_b & 0 \\
0 & 0 & \epsilon_c
\end{array}
\right),
\end{equation}
where $a$, $b$, and $c$ are the principal dielectric axes of the molecule.  Let
the $c$ axis represent the long axis of the molecule, while $a$ and $b$ are
orthogonal to that axis.  To transform this tensor into the laboratory
coordinate system, we make two rotations.  First, to represent the tilt of the
molecule with respect to the smectic layer normal, we rotate through the polar
angle $\eta$ about the molecular $b$ axis.  Second, to represent the
orientation of the tilt direction in the smectic layer plane, we rotate through
the azimuthal angle $\phi$ about the laboratory $z$ axis, the smectic layer
normal.  The result of these two rotation operations is
\begin{equation}
\epsilon=\left(
\begin{array}{c|c|c}
\begin{array}{l}
      \epsilon_a\cos^2\eta\cos^2\phi \\
      +\epsilon_c\sin^2\eta\cos^2\phi \\
      +\epsilon_b\sin^2\phi
\end{array} &
\begin{array}{l}
      -\epsilon_b\cos\phi\sin\phi \\
      +\epsilon_a\cos^2\eta\cos\phi\sin\phi \\
      +\epsilon_c\sin^2\eta\cos\phi\sin\phi
\end{array} &
(\epsilon_c-\epsilon_a)\cos\eta\sin\eta\cos\phi \\
\hline
\begin{array}{l}
      -\epsilon_b\cos\phi\sin\phi \\
      +\epsilon_a\cos^2\eta\cos\phi\sin\phi \\
      +\epsilon_c\sin^2\eta\cos\phi\sin\phi
\end{array} &
\begin{array}{l}
      \epsilon_a\cos^2\eta\sin^2\phi \\
      +\epsilon_c\sin^2\eta\sin^2\phi \\
      +\epsilon_b\cos^2\phi
\end{array} &
(\epsilon_c-\epsilon_a)\cos\eta\sin\eta\sin\phi \\
\hline
(\epsilon_c-\epsilon_a)\cos\eta\sin\eta\cos\phi &
(\epsilon_c-\epsilon_a)\cos\eta\sin\eta\sin\phi &
\epsilon_a\sin^2\eta+\epsilon_c\cos^2\eta
\end{array}
\right).
\label{rotatedtensor}
\end{equation}

We now make three assumptions about the distribution of molecular orientations.
First, we suppose that all molecules have the same value of the polar angle
$\eta$, which characterizes the tilt cone.  For simplicity, we suppose this
angle is independent of temperature and applied field.  Second, we suppose that
the molecules have a distribution of the azimuthal angle $\phi$.  In zero field
this distribution is uniform, but under an applied electric field $E$ (in the
$y$ direction) this distribution must be biased (in favor of tilt in the $x$
direction).  We assume the mean-field distribution function
\begin{equation}
\rho({\phi})=\rho_0\exp(EP_0\cos\phi/k_B T),
\label{distribution}
\end{equation}
where $\rho_0$ is a normalization factor, $T$ is the temperature, and $P_0$ is
an effective dipole moment coupling to the electric field, which will be
discussed further below.  Third, we suppose that there is no distribution of
rotations about the molecular long axes, i.e. that the molecules all have a
unique value of the third Euler angle.  This simplifying assumption is
justified by the idea that whatever microscopic interaction favors molecular
tilt must prefer a particular part of the molecule to point down toward the
smectic layers.  It implies that the molecular dipole moments are in the
smectic layer plane, tangent to the tilt cone.

Given these assumptions, we can average the dielectric
tensor~(\ref{rotatedtensor}) over the distribution
function~(\ref{distribution}).  The result is
\begin{equation}
\epsilon=\left(
\begin{array}{c|c|c}
\begin{array}{l}
 \epsilon_a\cos^2\eta
 \left(\frac{1}{2}+\frac{1}{2}\frac{I_2(EP_0/k_B T)}{I_0(EP_0/k_B T)}\right) \\
 +\epsilon_c\sin^2\eta
 \left(\frac{1}{2}+\frac{1}{2}\frac{I_2(EP_0/k_B T)}{I_0(EP_0/k_B T)}\right) \\
 +\epsilon_b
 \left(\frac{1}{2}-\frac{1}{2}\frac{I_2(EP_0/k_B T)}{I_0(EP_0/k_B T)}\right)
\end{array} &
0 &
(\epsilon_c-\epsilon_a)\cos\eta\sin\eta
\frac{I_1(EP_0/k_B T)}{I_0(EP_0/k_B T)} \\
\hline
0 &
\begin{array}{l}
 \epsilon_a\cos^2\eta
 \left(\frac{1}{2}-\frac{1}{2}\frac{I_2(EP_0/k_B T)}{I_0(EP_0/k_B T)}\right) \\
 +\epsilon_c\sin^2\eta
 \left(\frac{1}{2}-\frac{1}{2}\frac{I_2(EP_0/k_B T)}{I_0(EP_0/k_B T)}\right) \\
 +\epsilon_b
 \left(\frac{1}{2}+\frac{1}{2}\frac{I_2(EP_0/k_B T)}{I_0(EP_0/k_B T)}\right)
\end{array} &
0 \\
\hline
(\epsilon_c-\epsilon_a)\cos\eta\sin\eta
\frac{I_1(EP_0/k_B T)}{I_0(EP_0/k_B T)} &
0 &
\epsilon_a\sin^2\eta+\epsilon_c\cos^2\eta
\end{array}
\right),
\label{averagetensor}
\end{equation}
where $I_0$, $I_1$, and $I_2$ are the modified Bessel functions.

To model the experimental results, we must predict the optical properties of a
sample for light propagating in the $y$ direction, parallel to the applied
electric field.  For that reason, we diagonalize the average dielectric tensor
in the $xz$ plane.  The eigenvectors give the principal optical axes of the
sample.  In particular, the optical tilt $\theta(E)$ is the angle between the
eigenvectors and the $x$ and $z$ axes.  The eigenvalues give the dielectric
constants along the principal optical axes.  The indices of refraction are the
square roots of these dielectric constants, and the birefringence is then the
difference between these square roots.

This diagonalization can be done exactly in the two limiting cases of low field
and high field.  For $E \to 0$, the tensor is already diagonal, and we obtain
\begin{mathletters}
\label{lowfieldlimit}
\begin{eqnarray}
& \theta(0)=0, \\
& \Delta n(0)=\sqrt{\epsilon_a\sin^2\eta+\epsilon_c\cos^2\eta}
-\sqrt{\frac{\epsilon_a\cos^2\eta+\epsilon_c\sin^2\eta+\epsilon_b}{2}} .
\end{eqnarray}
\end{mathletters}
By comparison, for $E \to\infty$, diagonalization gives
\begin{mathletters}
\label{highfieldlimit}
\begin{eqnarray}
& \theta(\infty)=\eta, \\
& \Delta n(\infty)=\sqrt{\epsilon_c}-\sqrt{\epsilon_a} .
\end{eqnarray}
\end{mathletters}
Note that the high-field limit shows that maximum possible birefringence, which
comes from the difference between the dielectric constant $\epsilon_c$ along
the long axis of the molecule and the dielectric constant $\epsilon_a$
perpendicular to the long axis.  The low-field limit shows a lower
birefringence, because it mixes the dielectric components in a rotational
average.

For intermediate values of the electric field, we diagonalize the tensor
numerically using Mathematica.  This numerical procedure shows that the
predicted birefringence and tilt have the same general form as the experimental
data.  For low fields, the tilt increases linearly and the birefringence
increases quadratically with field.  They both saturate around a field of
$k_B T/P_0$ and approach a limiting value at high field.  The question is thus
how well the prediction can fit the data for birefringence and tilt
{\it simultaneously.}

Before we go on to the fits, we should briefly discuss the interpretation of
the parameter $P_0$.  In the mean-field distribution function of
Eq.~(\ref{distribution}), $P_0$ is the effective dipole moment that couples to
the applied electric field.  Because the molecules undergo orientational
fluctuations in large correlated groups, $P_0$ can be much greater than the
dipole moment of a single molecule.  Near a second-order transition from the
SmA to the SmC phase, it should increase as a power law.  Because $P_0$
represents the susceptibility of the tilt angle to an applied electric field,
it should scale with the susceptibility exponent $\gamma$,
\begin{equation}
P_0(T) \propto (T-T_{AC})^{-\gamma}.
\end{equation}
The SmA--SmC transition should be in the universality class of the
three-dimensional $xy$ model, and hence we expect
$\gamma\approx1.33$~\cite{degennes93}.  This expected scaling will be tested by
the fits in the following section.

\section{Fitting}

To compare the theory with the experimental data, we note that the theory
involves five parameters:  the cone angle $\eta$, the dielectric parameters
$\epsilon_a$, $\epsilon_b$, and $\epsilon_c$, and the effective dipole moment
$P_0$.  The first four of these parameters should be independent of temperature
and should depend only on the liquid-crystalline material, while the last
parameter $P_0$ should be a function of temperature.

To determine the cone angle $\eta$, we use the limiting value of the tilt data
at high field, following Eq.~(\ref{highfieldlimit}a).  We use the
lowest-temperature data set because it has the clearest features.  To determine
the dielectric parameters $\epsilon_a$, $\epsilon_b$, and $\epsilon_c$, we use
the limiting values of the birefringence data at low and high fields, again
using the lowest-temperature data set.  Equations~(\ref{lowfieldlimit}b)
and~(\ref{highfieldlimit}b) then give two constraints on the three dielectric
parameters.  For a third constraint, we assume that the isotropically averaged
index of refraction $\sqrt{(\epsilon_a+\epsilon_b+\epsilon_c)/3}=1.6$.  This
value of 1.6 is just a typical value for an organic liquid, and we have
confirmed that the results are not sensitive to this particular choice.  With
these three constraints, we can solve for $\epsilon_a$, $\epsilon_b$, and
$\epsilon_c$.  The results for all the temperature-independent parameters are
listed in Table I.  Note that the cone angles are very similar, $33^\circ$ in
DSiKN65 and $34^\circ$ in TSiKN65, and the dielectric parameters are also quite
similar between the liquid crystals.  Presumably this is because of the
chemical similarity between these two materials.

Once those parameters are determined, there is only one remaining
temperature-dependent fitting parameter $P_0(T)$.  To determine this parameter,
we fit the combined data for tilt vs.\ field and birefringence vs.\ field at
each temperature.  In this fit, we must combine the two contributions to
$\chi^2$ with appropriate weighting factors.  A reasonable choice is to weight
the birefringence data (unitless) by a factor of 1000 relative to the tilt data
(in radians), which gives equally good fits to both data sets.  The fits are
shown by the solid lines in Figs.~2(a-b) and~3(a-b), and the extracted values
of $P_0(T)$ are listed in Table I.  Clearly the theory captures the field
dependence of the tilt and birefringence data.  The fits are qualitatively good
for all of the data and quantitatively good for most of the data.

An alternative way to look at the data is to plot the birefringence vs.\ tilt
angle.  As mentioned in Sec.~II, the data at all temperatures collapse onto a
single universal curve for each material.  This data collapse is indeed a
feature of the theory:  Because the average dielectric tensor of
Eq.~(\ref{averagetensor}) depends on field and temperature only through the
combination $E P_0(T)/k_B T$, the theory predicts a universal curve that
depends only on $\eta$, $\epsilon_a$, $\epsilon_b$, and $\epsilon_c$.  In
Figs.~2(c) and~3(c), we plot the theoretical curve along with the data.  Note
that the {\it endpoints\/} of this curve are fixed by the fit parameters, but
the shape of the curve between the endpoints is determined by the theory with
no further choice of parameters.  This shape is generally close to the data,
although there is some clear discrepancy.

We have tried slightly different estimates for the zero-field and high-field
limits of the tilt and birefringence, as well as a different fitting procedure
that determines all the parameters from the birefringence data and then uses
them to calculate the tilt angle.  The results of all these variations are
quite similar to what is shown here.  The differences between the theoretical
curves and the data are always present at about the same level.

Note that these fits imply that the molecules are biaxial, with
$\epsilon_a\not=\epsilon_b\not=\epsilon_c$.  For comparison, we considered a
uniaxial model with $\epsilon_a=\epsilon_b\not=\epsilon_c$.  This model gives
good fits to the birefringence data, but it implies a cone angle $\eta$ of
24--26$^\circ$, which is less than the observed tilt angle.  As a result, the
fits involving the tilt angle ($\theta$ vs.\ $E$ and $\Delta n$ vs.\ $\theta$)
are unsatisfactory.  (This inconsistency occurs even if we eliminate the
constraint on the isotropically averaged index of refraction.)

In Sec.~III, we argued that the value of $P_0(T)$ should increase as the
temperature decreases toward the SmA-SmC transition.  The fit results in
Table~I are consistent with this trend.  To analyze the temperature dependence,
we plot $P_0$ vs.~$T$ in Fig.~4(a-b) and fit the data to the power law
\begin{equation}
P_0(T)=A\left(\frac{T-T_{AC}}{T_{AC}}\right)^{-\gamma}.
\end{equation}
The power law gives a very good fit to the observed temperature dependence,
with the fitting parameters listed in Table~II.  Note that the exponent
$\gamma$ is 1.51 for DSiKN65 and 1.75 for TSiKN65.  This exponent is somewhat
larger than the expected value of 1.33, but we do not have enough data close to
the transition to determine the exponent precisely.  Overall, the fitting
results are consistent with the theoretical concept that $P_0$ is an effective
dipole moment that grows larger as the system approaches the SmA-SmC
transition, following a power-law scaling relation.

A further consistency check comes from the amplitude of the power-law
variation.  The amplitude $A$ is 0.54 debye in DSiKN65 and 0.44 debye in
TSiKN65, where 1 debye $=10^{-18}$ esu cm.  This is the same order of magnitude
as a typical molecular dipole moment of 1--2 debye~\cite{goodby91}.  Over the
experimental temperature range, $P_0(T)$ increases from roughly $10^2$ to
$10^3$ times this value.

One aspect of the fitting results for $P_0(T)$ is surprising.  Experimentally,
the SmA-SmC transition occurs within 0.5~$^\circ$C of the lowest temperature
for which tilt angle and birefringence were measured.  However, the fits for
$P_0(T)$ shown in Table II indicate a second-order transition temperature
almost 2~$^\circ$C below the actual transition temperature.  Power-law fits to
the tilt angle and birefringence data vs.\ temperature at the lowest nonzero
value of the electric field also indicate second-order transition temperatures
consistent with those in Table II.

One possible explanation for the difference between the experimental and the
fit transition temperatures is that the transition is weakly first-order, with
a small discontinuous change in the tilt angle and birefringence.  To test this
possibility, we looked for hysteresis upon heating and cooling through the
transition in DSiKN65 using a differential scanning calorimeter.  The
transition always occurred at a higher temperature upon heating as opposed to
cooling.  When this temperature difference was plotted versus the
heating/cooling rate (0.02 to 0.30~$^\circ$C/min), it extrapolated linearly to
0.05 $^\circ$C at zero heating/cooling rate.  This hysteresis indicates that
the transition has a slight first-order character.

Another possible explanation for this difference is that there is another phase
between the SmC and SmA phases.  Since this transition involves the
establishment of long-range azimuthal order of the tilt, there could be an
intermediate phase, perhaps one with a discrete distribution of azimuthal
angles.  We see no evidence for this in the optical and DSC data, but these
types of measurements may be insensitive to such structural changes.

\section{Discussion}

In this paper, we have presented a theory for the orientational distribution of
molecules in de Vries-type SmA liquid crystals.  This theory makes the simplest
possible assumptions about the distribution of molecular orientations on a tilt
cone, and gives predictions for the dependence of tilt angle and
birefringence on electric field and temperature.  We have compared these
predictions with experimental data for the tilt and birefringence near the
SmA-SmC transition in the two materials DSiKN65 and TSiKN65.  The overall
quality of the fits is good, considering that we are fitting a simple model to
a large amount of data over a wide range of electric field and reduced
temperature.  Furthermore, the fits give quantitatively reasonable values for
the effective dipole moment, and show how this quantity increases as the system
approaches the SmA-SmC transition.

While the agreement between the theory and the experiment is generally good,
there are clearly some deviations.  These deviations show that the experimental
system has some behavior that is more complex than the simple assumptions of
the theory.  First, the cone angle probably has some dependence on temperature
and electric field. This dependence is shown by measurements of the layer
spacing in TSiKN65~\cite{spector00}:  For the range of electric field and
temperature that we have studied, the layer spacing changes by roughly 0.6\%
with field and roughly 0.1\% with temperature.  Second, the molecules may have
a distribution of rotations about the molecular long axes, i.e. a distribution
of dipole moment orientations relative to the tilt cone, and this distribution
may change as a function of field and temperature.  Third, the system may have
a distribution of molecular conformations, and this distribution may also
change with field and temperature.  We have not considered these effects in our
current theory, because we wish to explain the main trends in the data with the
simplest possible theory and to avoid adding further fitting parameters.
However, these effects can be studied in future work.

As a final point, we speculate that there are not really two separate classes
of SmA liquid crystals:  ``conventional'' and de Vries-type.  Rather, there may
be a whole spectrum of materials between these two extremes.  On one end of the
spectrum are SmA liquid crystals with a very small cone angle.  When an
electric field is applied, the main response is that the molecules tilt
uniformly by much more than the cone angle.  These are the ``conventional'' SmA
materials that undergo layer contraction.  On the other end of the spectrum are
SmA liquid crystals with large cone angles.  As an electric field is applied,
the main response is the establishment of long-range azimuthal order, with a
relatively small change in the magnitude of the cone angle.  These are the SmA
materials that tilt with extremely little layer contraction.  In between these
limiting cases, other liquid crystals may undergo substantial changes in
{\it both\/} the cone angle {\it and\/} the azimuthal distribution in response
to an electric field. The materials that we have studied, DSiKN65 and TSiKN65,
are clearly near the de Vries limit of this spectrum, but their response to an
electric field gives insight into the full range of behavior that is possible
in the SmA phase.

\acknowledgments

We would like to thank J.~Naciri for synthesizing the liquid crystals,
B.~R.~Ratna and M.~S.~Spector for helpful discussions, and B.~T.~Weslowski for
assistance with the experiments.  We particularly thank F.~J.~Bartoli,
S.~R.~Flom, and J.~R.~Lindle for their collaboration on an earlier part of this
study, published in Ref.~\cite{lindle99}.  This research was supported by the
Office of Naval Research and the Naval Research Laboratory.

\epsfclipon

\begin{figure}
\centering\leavevmode\epsfxsize=3.375in\epsfbox{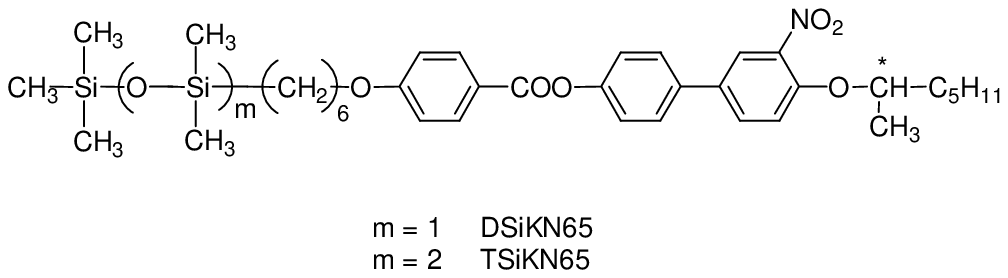}\bigskip
\caption{Molecular structure of the siloxane-containing liquid crystals studied
in this paper.}
\end{figure}

\begin{figure}
\vbox{
\noindent \centering\leavevmode\epsfxsize=3.375in\epsfbox{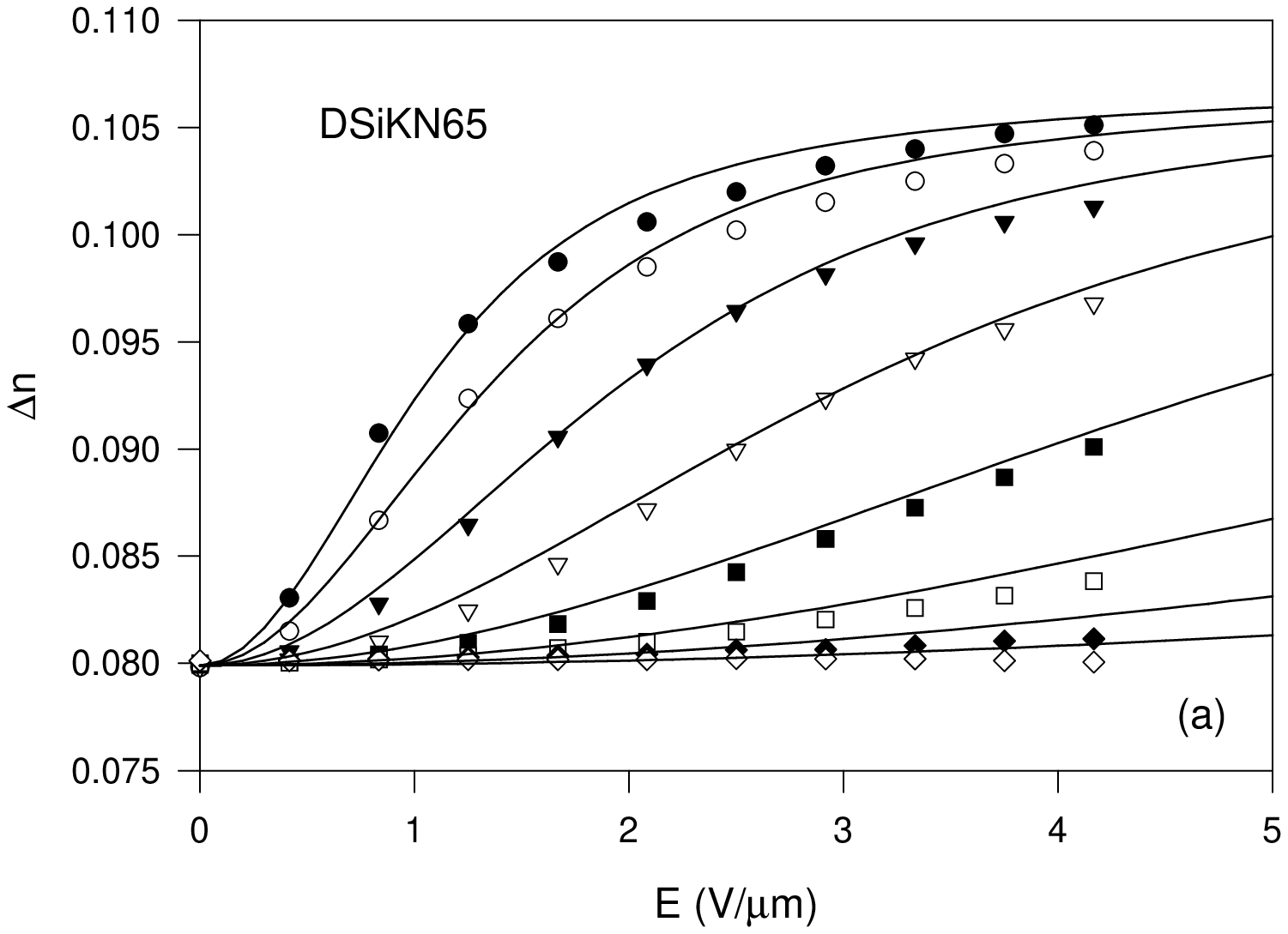}\bigskip

\centering\leavevmode\epsfxsize=3.375in\epsfbox{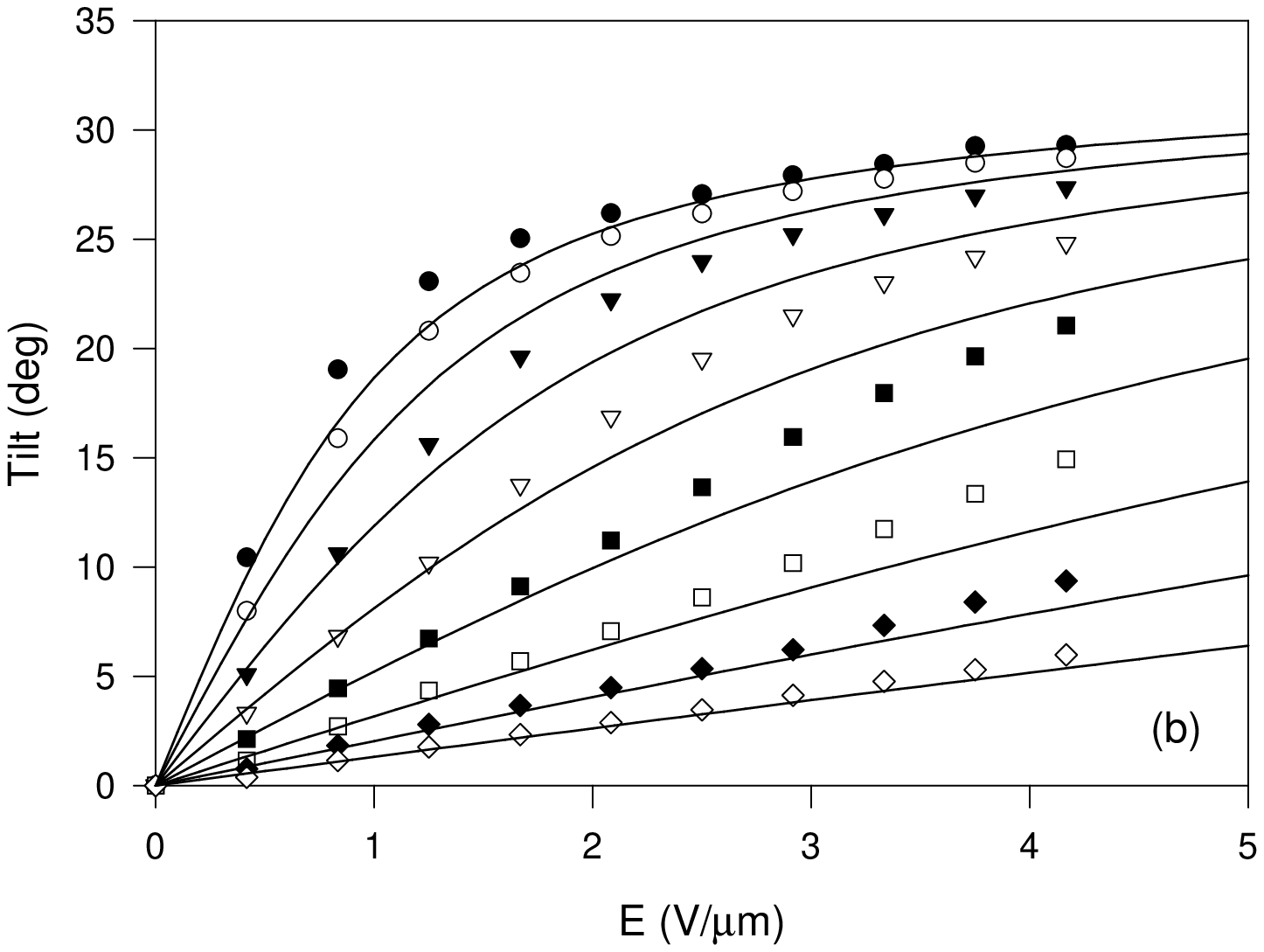}\bigskip

\centering\leavevmode\epsfxsize=3.375in\epsfbox{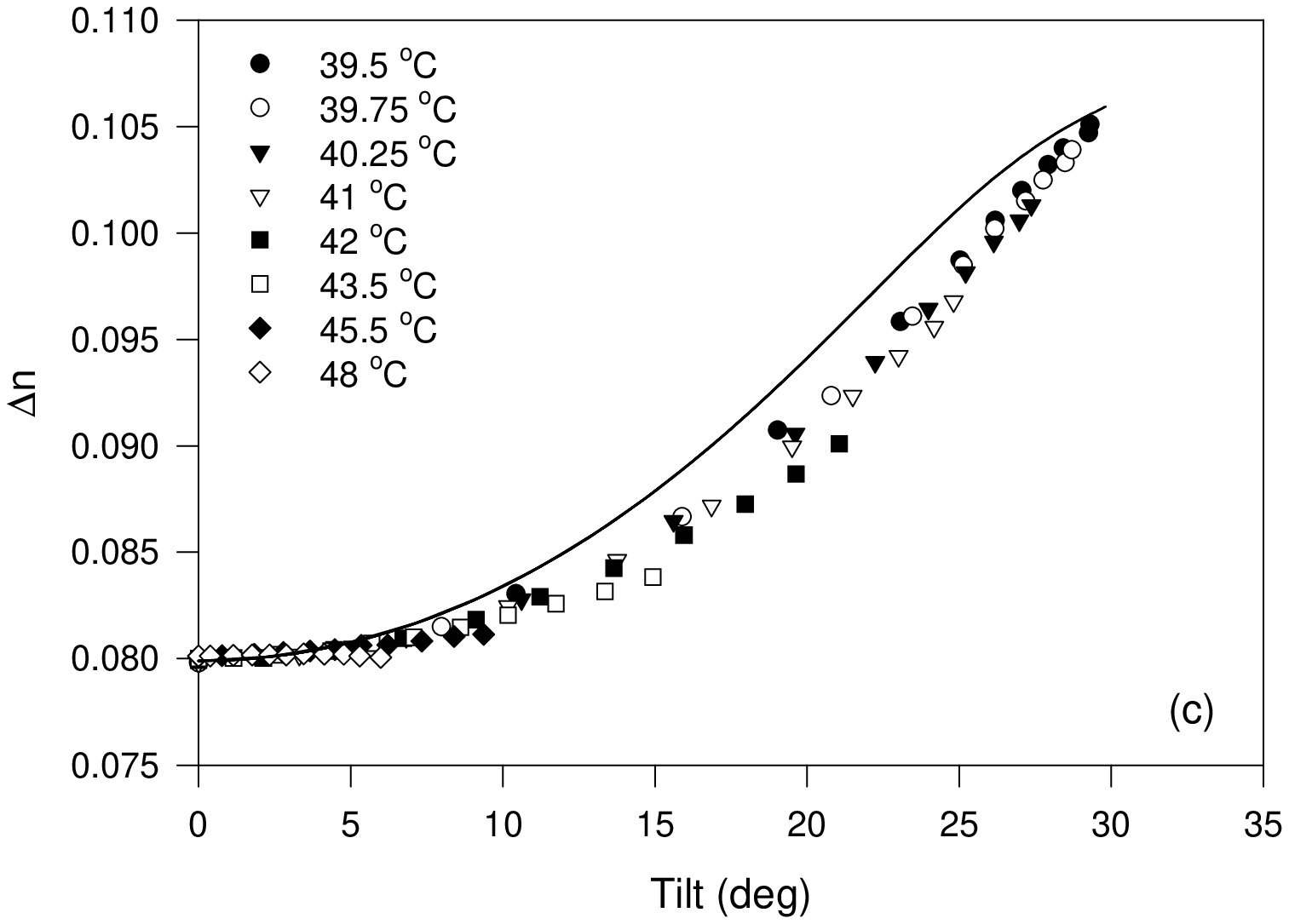}\bigskip
\caption{Symbols---Data for the field-dependent birefringence and tilt of
DSiKN65 at several temperatures:  39.5, 39.75, 40.25, 41, 42, 43.5, 45.5, and
48~$^\circ$C (top to bottom).  Lines---Fits for the field-dependent
birefringence and tilt at the same temperatures (top to bottom).
(a)~Birefringence vs. field.  (b)~Tilt vs. field.  (c)~Birefringence vs. tilt.}
}
\end{figure}

\begin{figure}
\vbox{
\noindent \centering\leavevmode\epsfxsize=3.375in\epsfbox{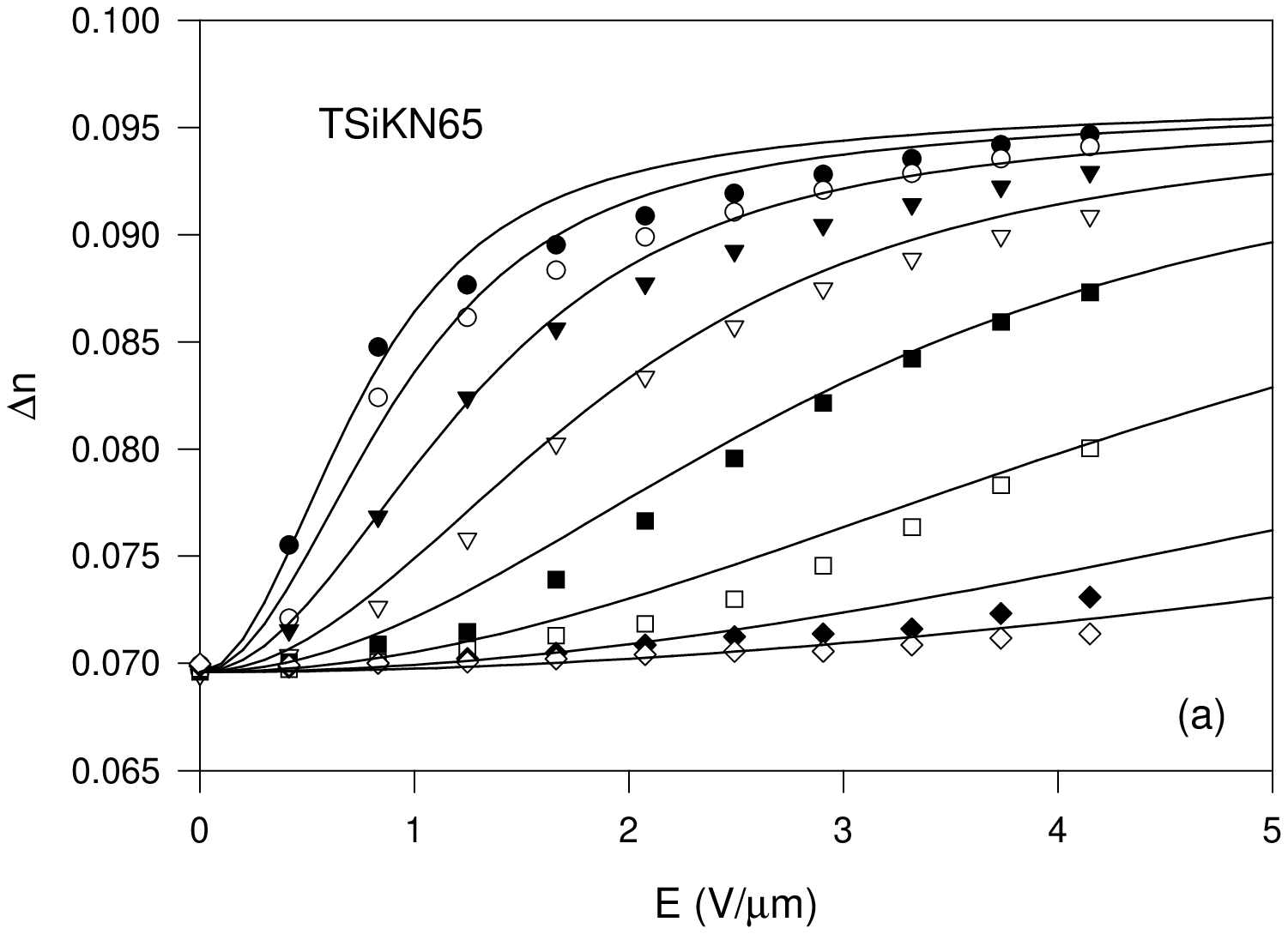}\bigskip

\centering\leavevmode\epsfxsize=3.375in\epsfbox{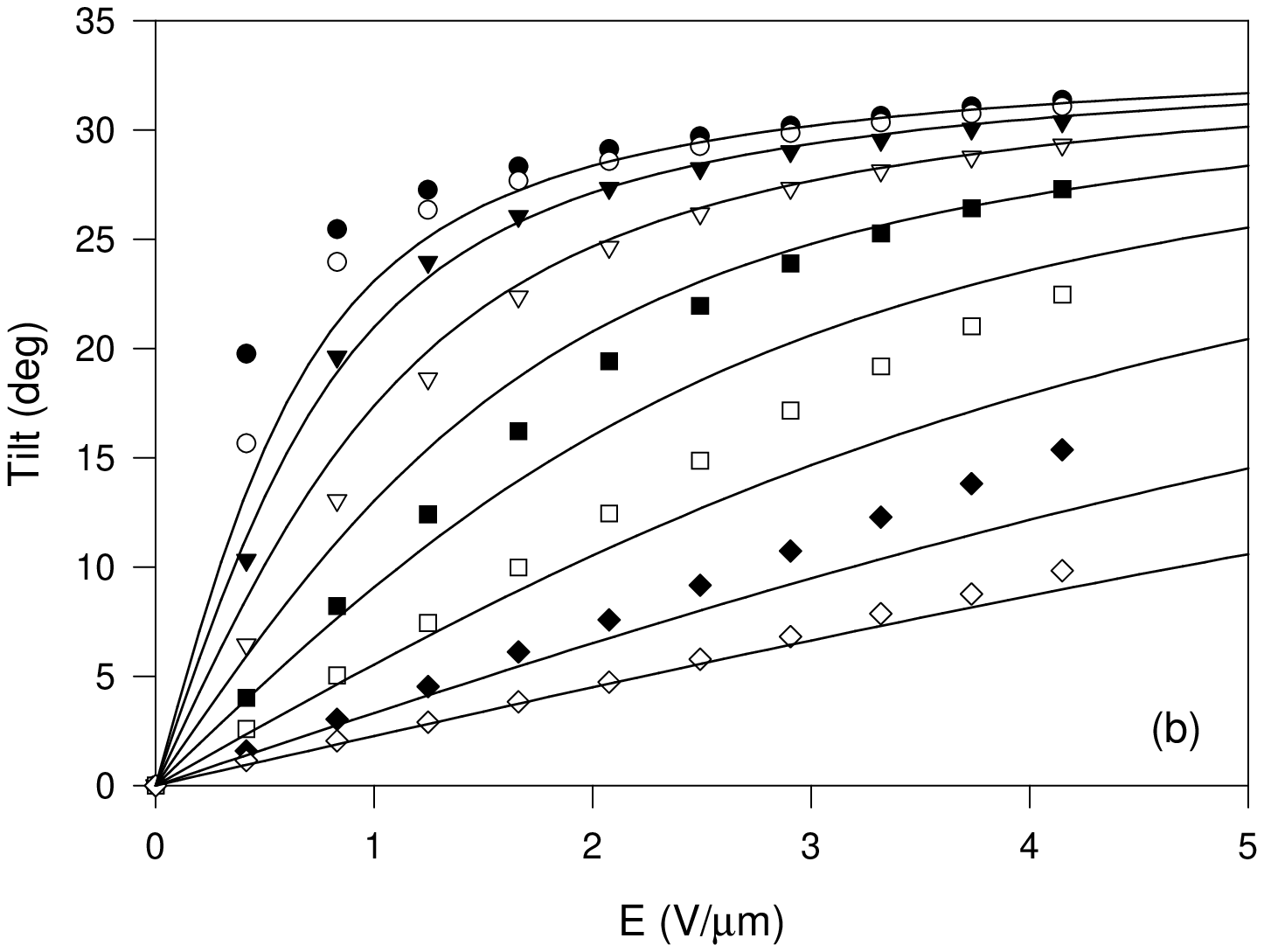}\bigskip

\centering\leavevmode\epsfxsize=3.375in\epsfbox{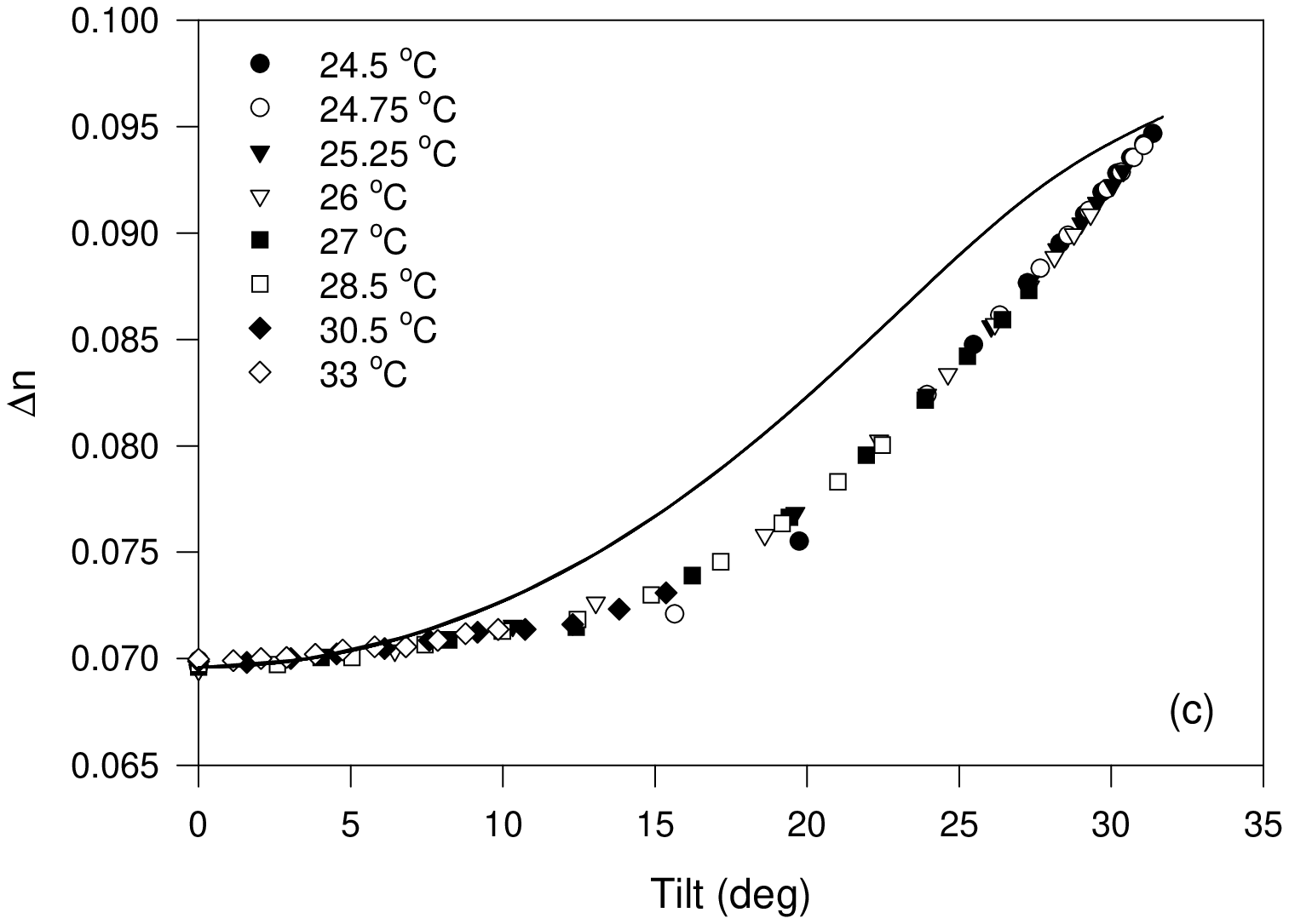}\bigskip
\caption{Symbols---Data for the field-dependent birefringence and tilt of
TSiKN65 at several temperatures:  24.5, 24.75, 25.25, 26, 27, 28.5, 30.5, and
33~$^\circ$C (top to bottom).  Lines---Fits for the field-dependent
birefringence and tilt at the same temperatures (top to bottom).
(a)~Birefringence vs. field.  (b)~Tilt vs. field.  (c)~Birefringence vs. tilt.}
}
\end{figure}

\begin{figure}
\vbox{
\noindent \centering\leavevmode\epsfxsize=3.375in\epsfbox{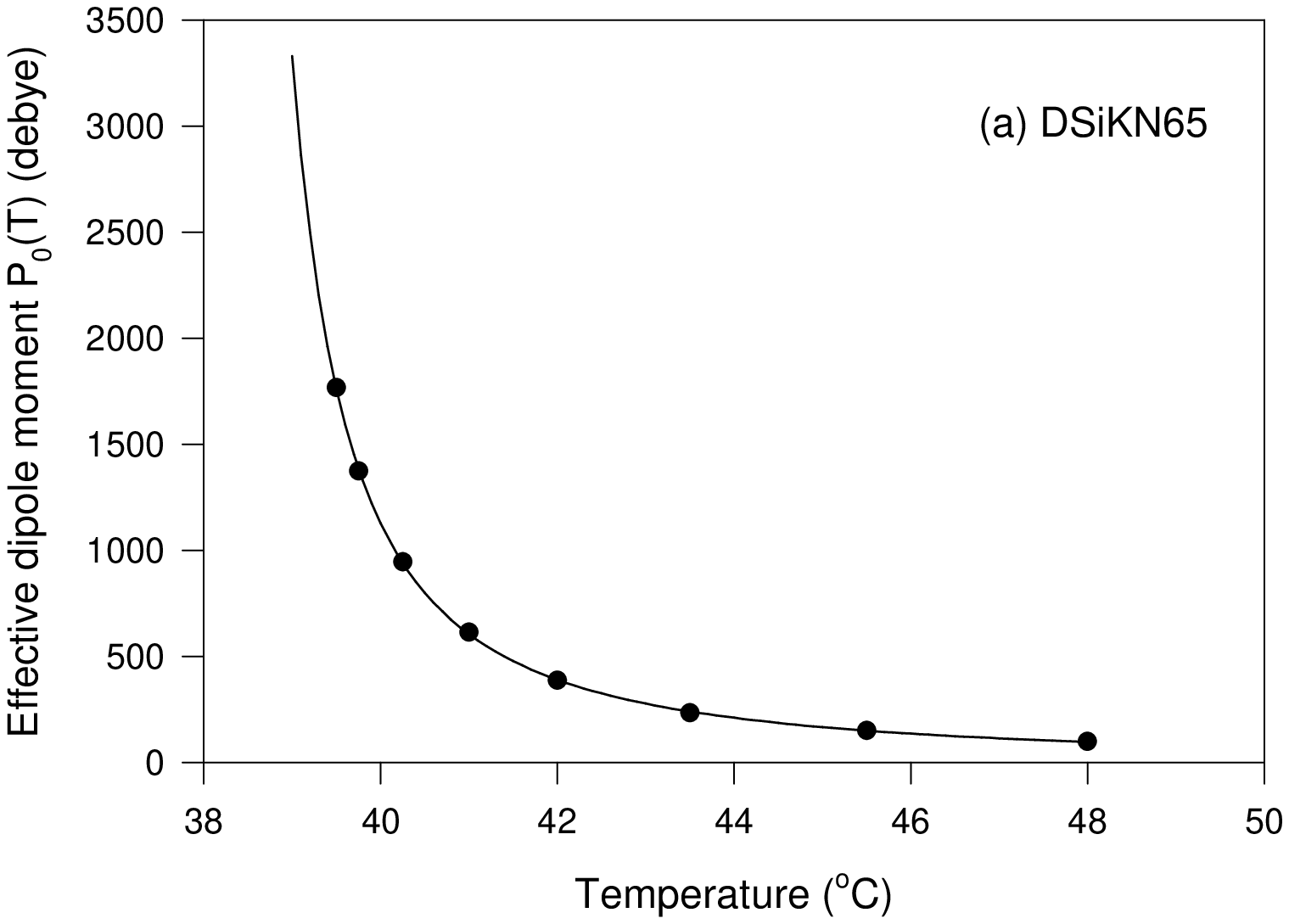}\bigskip

\centering\leavevmode\epsfxsize=3.375in\epsfbox{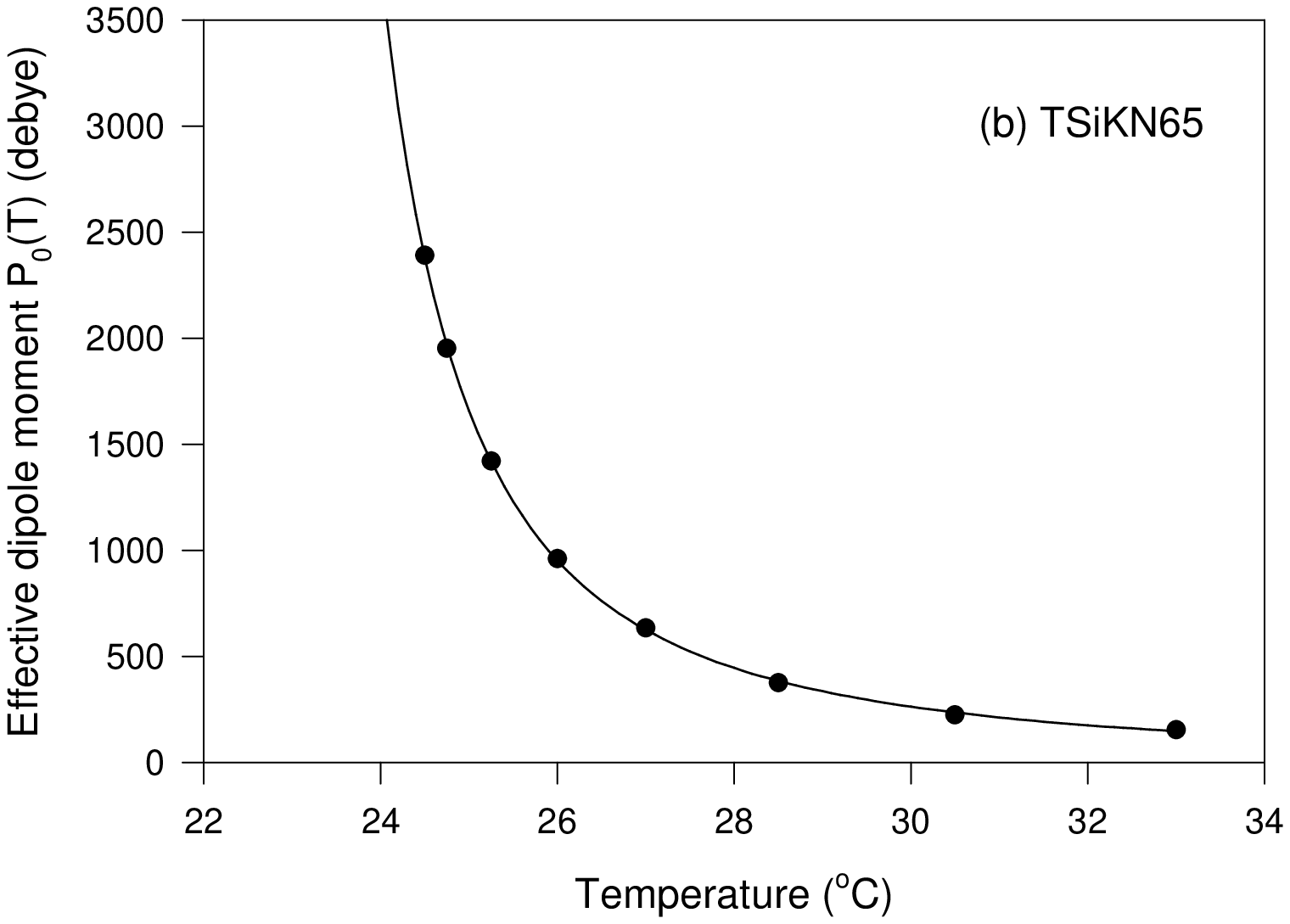}\bigskip
\caption{Symbols---Effective dipole moment $P_0(T)$, from Table~I.
Lines---Power-law fits for the temperature dependence of $P_0(T)$.
(a)~DSiKN65.  (b)~TSiKN65.}
}
\end{figure}

\begin{table}
\caption{Fit parameters for the two liquid crystals studied.  The first four
parameters are temperature-independent, while $P_0$ depends on temperature.}
\begin{tabular}{cdcdc}
Parameter   &Temp~($^\circ$C)&DSiKN65   &Temp~($^\circ$C)&TSiKN65   \\
\tableline
$\eta$      &                &$33^\circ$&                &$34^\circ$\\
$\epsilon_a$&                &2.484     &                &2.493     \\
$\epsilon_b$&                &2.360     &                &2.379     \\
$\epsilon_c$&                &2.836     &                &2.808     \\
\tableline
$P_0(T)$    & 39.5           &1768      & 24.5           &2390      \\
(debye)     &39.75           &1373      &24.75           &1952      \\
            &40.25           & 946      &25.25           &1420      \\
            & 41.0           & 614      & 26.0           & 961      \\
            & 42.0           & 386      & 27.0           & 633      \\
            & 43.5           & 233      & 28.5           & 376      \\
            & 45.5           & 150      & 30.5           & 224      \\
            & 48.0           &  97      & 33.0           & 154
\end{tabular}
\end{table}

\begin{table}
\caption{Power-law fit parameters for the temperature dependence of the
$P_0(T)$ data in Table I.}
\begin{tabular}{ccc}
Parameter           &DSiKN65      &TSiKN65      \\
\tableline
$\gamma$            &$1.51\pm0.06$&$1.75\pm0.08$\\
$T_{AC}$ ($^\circ$C)&$38.0\pm0.1$ &$22.3\pm0.1$ \\
$A$ (debye)         &$0.54\pm0.13$&$0.44\pm0.12$
\end{tabular}
\end{table}

\end{document}